%% file: main.tex
\documentclass[conference]{IEEEtran}
\IEEEoverridecommandlockouts
\usepackage{cite}
\usepackage{amsmath,amssymb,amsfonts}
\usepackage{graphicx}
\usepackage{textcomp}
\usepackage{floatrow}
\usepackage{xcolor}
\usepackage{tikz}
\usepackage{pgfplots}
\pgfplotsset{compat=1.5}
\usepackage{subcaption}
\usepackage{microtype}
\usepackage{graphicx}

\usepackage{dingbat}
\usepackage{adjustbox}
\usepackage{booktabs} 
\usepackage{dsfont}
\usepackage{hyperref}

\usepackage{amsmath}
\usepackage{amssymb}
\usepackage{mathtools}
\usepackage{amsthm}
\usepackage{algorithm,algorithmic}
\usepackage[capitalize,noabbrev]{cleveref}
\usetikzlibrary{backgrounds,arrows,shapes.geometric,shapes.misc,positioning,patterns, bayesnet}
\usepgfplotslibrary{fillbetween}
\usepackage{tikzsymbols}
\tikzstyle{block}=[draw opacity=0.7,line width=1.4cm]

\theoremstyle{plain}
\newtheorem{theorem}{Theorem}[section]
\newtheorem{proposition}[theorem]{Proposition}

\theoremstyle{definition}
\newtheorem{definition}[theorem]{Definition}

\theoremstyle{remark}

\usepackage[textsize=tiny]{todonotes}

\makeatletter
\newcommand{\linebreakand}{%
  \end{@IEEEauthorhalign}
  \hfill\mbox{}\par
  \mbox{}\hfill\begin{@IEEEauthorhalign}
}
\makeatother

\def\BibTeX{{\rm B\kern-.05em{\sc i\kern-.025em b}\kern-.08em
    T\kern-.1667em\lower.7ex\hbox{E}\kern-.125emX}}
\begin{document}

\title{Efficient Liquidity Providing via Margin Liquidity}
\author{\IEEEauthorblockN{Yeonwoo Jeong$^{*}$}
\IEEEauthorblockA{\textit{Dept. Computer Science and Engineering,} \\
\textit{Seoul National University}\\
yeonwoo@mllab.snu.ac.kr}
\and
\IEEEauthorblockN{Chanyoung Jeoung$^{*}$}
\IEEEauthorblockA{\textit{School of Medicine,}\\ \textit{CHA University} \\
jy20140109@snu.ac.kr}
\and
\IEEEauthorblockN{Hosan Jeong}
\IEEEauthorblockA{\textit{Dept. Computer Science and Engineering,} \\
\textit{Seoul National University}\\
nei0964@snu.ac.kr}
\linebreakand
\IEEEauthorblockN{ SangYoon Han}
\IEEEauthorblockA{\textit{Ministry of Justice, ROK}\\
samsyhan@snu.ac.kr}
\and
\IEEEauthorblockN{Juntae Kim}
\IEEEauthorblockA{\textit{Formula Labs}\\
juntaegim99@gmail.com}
}

\maketitle

\begin{abstract}
\input{text/01abstract.tex}
\end{abstract}
%\blfootnote{$^*$Equal contribution.}

\section{Introduction}
\input{text/02intro.tex}

\section{Preliminary: Uniswap V2}
\input{text/03preliminary.tex}

\section{Method}
\input{text/04method.tex}

\section{Related works}
\input{text/05relatedworks}

\section{Application}
\label{sec:app}
\input{text/06application.tex}

\section{Experiments}
\input{table/result}
\input{text/07experiments}

\section{Conclusion}
\input{text/08conclusion}

\bibliographystyle{IEEETranS}
\bibliography{refs}

\end{document}

%% file: text/01abstract.tex
The limit order book mechanism has been the core trading mechanism of the modern financial market. In the cryptocurrency market, centralized exchanges also adopt this limit order book mechanism and a centralized matching engine dynamically connects the traders to the orders of market makers. Recently, decentralized exchanges have been introduced and received considerable attention in the cryptocurrency community. A decentralized exchange typically adopts an automated market maker, which algorithmically arbitrates the trades between liquidity providers and traders through a pool of crypto assets. Meanwhile, the liquidity of the exchange is the most important factor when traders choose an exchange. However, the amount of liquidity provided by the liquidity providers in decentralized exchanges is insufficient when compared to centralized exchanges. This is because the liquidity providers in decentralized exchanges suffer from the risk of divergence loss inherent to the automated market making system. To this end, we introduce a new concept called margin liquidity and leverage this concept to propose a highly profitable margin liquidity-providing position. Then, we extend this margin liquidity-providing position to a virtual margin liquidity-providing position to alleviate the risk of divergence loss for the liquidity providers and encourage them to provide more liquidity to the pool. Furthermore, we introduce a representative strategy for the margin liquidity-providing position and backtest the strategy with historical data from the BTC/ETH market. Our strategy outperforms a simple holding baseline. We also show that our proposed margin liquidity is 8K times more capital efficient than the concentrated liquidity proposed in Uniswap V3.

%% file: text/02intro.tex
The cryptocurrency market has flourished remarkably over the recent years, with the daily trading volume to market cap ratio reaching $4\times$ higher than the stock market\footnote{The global cryptocurrency market cap is \$ 1.26T and the daily trading volume in the cryptocurrency market is at \$ 65.9B (2022-06-11, Coingecko.com) while the U.S. stock market cap is \$43.8T and daily trading volume in the U.S. stock market is \$ 534B (2022-06-13, Nasdaqtrader).}. At the center of the cryptocurrency market lies centralized exchange (CEX), which adopts the traditional limit order book (LOB) mechanism that provides cryptocurrency trading with a centralized matching engine \cite{nasdaq}. Market makers quote prices in a LOB, and the centralized matching engine dynamically connects the traders who want to buy or sell to the orders of market makers in the book, as illustrated in \Cref{fig:relationCEX}. On the other hand, a new trading mechanism called automated market makers (AMM) has been proposed with the advent of decentralized exchange (DEX), which provides direct cryptocurrency tradings through the pools of crypto assets instead of the centralized matching engine in CEX \cite{uniswap, uniswapv3, balancer, curvev1, curvev2, sokamm, dexamm, dynamic, malamud2017decentralized,capponi2021adoption,han2021trust, sokdefi, defiattack}. In this paper, we denote a DEX that adopts AMM as DEX-AMM. AMM algorithmically arbitrates the trades between liquidity providers (LPs), who are market makers in DEX, and traders. LPs deposit their assets in the pools operated by AMM and traders swap their assets in the pools. Through the AMM, LPs receive swap fees from traders in return for providing liquidity, as shown in \Cref{fig:relationDEX}\cite{feereport}.

% Note, the trading volume of DEX has dramatically increased to occupy more than 10\% of the trading volume of CEX. 
% Even though DEX featuring transparency and trustworthiness gains investors’ trust \cite{}, traditional traders prefer CEX to DEX due to the following reasons: relatively low transaction fee in CEX compared to heavy gas fee in DEX and fast transaction speed of CEX. However, since liquidity is hidden in LOB, CEX is vulnerable to possible market manipulation and cascading liquidation. 
\input{fig/releation.tex}

Liquidity in the exchange is one of the most significant factor when traders choose an exchange since traders can trade whenever they want only if the liquidity is sufficient \cite{barbon2021quality, aoyagi2021coexisting}. On the contrary, if the liquidity provided is insufficient, traders might experience an unexpectedly huge price impact or have to wait until a reasonable price comes. Liquidity in DEX-AMM is determined by the amount of liquidity the LPs provide. However, in current DEX-AMMs, the liquidity is not sufficiently provided by the LPs due to the risk of divergence loss inherent to the AMM system \cite{capponi2021adoption}. Divergence loss happens when the ratio of tokens deposited in the pool changes. We will describe the details of the divergence loss in \Cref{divloss}. To solve this problem, we introduce a new concept of liquidity, \emph{margin liquidity}, and propose a margin liquidity-providing position that is highly profitable and can be applied to both CEX and DEX-AMM. Then, we extend this margin liquidity-providing position to a \emph{virtual} margin liquidity-providing position to alleviate the risk of divergence loss of LPs in DEX-AMM and encourage them to provide more liquidity to the pool. Furthermore, we introduce a representative strategy for the margin liquidity-providing position and evaluate the strategy by backtesting it on the historical data of the ETH/BTC market with the Sharpe ratio\cite{sharperatio}, maximum drawdown (MDD)\cite{mdd, drawdown}, and the rate of return (ROR)\cite{ror}. We also show that our proposed margin liquidity is much more capital efficient than the concentrated liquidity of Uniswap V3 \cite{uniswapv3}. 

%We expect our approaches will complement LOB in the modern financial market.

%Otherwise, 

%Price impact in the LOB system is the endogenous outcome of the interactions between traders and market makers. 

%Trading volume is directly related to the revenue of Exchange. In the trend market, traders take either long or short positions and one of the positions takes a large profit against the other position. Therefore, the chance to earn large profits lures many market participants. However, in the sideways markets, the chance to earn large profit decreases and trading volume decreases as follows. Margin liquidity market maker Exchange fosters LPs to open margin liquidity providing position and close margin liquidity providing position even in sideways markets. Then, Exchange can make a profit even in sideways markets.

%Therefore, informed LPs withdraw their assets in the DEX pool when a volatile market is expected. On the other hand, noise LPs are likely to lose their assets by the divergence loss in the volatile market even though they are rewarded by high trading fees. In the debacle of LUNA, many LPs faced a huge loss of assets. In the volatile crypto market, LPs should frequently repeat withdrawals and deposits to maximize their profits while deviating from the divergence loss. 

%% file: fig/releation.tex
\begin{figure*}
\footnotesize
\def\W{1}     %width
\def\H{1}     %height
\centering %
\begin{subfigure}[t]{0.26\linewidth}
\centering
\begin{tikzpicture}[state/.style={circle, draw, minimum size=1.2cm}]
    \node[state] at (0,0) (x) {\textbf{Trader}};%
    \node[state, right=of x, red]  at (\W, 0) (z) {\textbf{MM}}; %
    \path (x) edge[ bend left=40,->] (z); 
    \path (z) edge[ bend left=40,->] (x); 
    \node[]at (1.4*\W,\H){buy or sell};
    \node[]at (1.4*\W,-\H){liquidity};
\end{tikzpicture}
\caption{Trader and market maker (MM)}
\label{fig:relationCEX}
\end{subfigure}\hfill
\begin{subfigure}[t]{0.26\linewidth}
\centering
\begin{tikzpicture}[state/.style={circle, draw, minimum size=1.2cm}]
    \node[state] at (0,0) (x) {\textbf{Trader}};%
    \node[state, right=of x, blue] at (\W, 0)(z) {\textbf{LP}}; %
    \path (x) edge[ bend left=40,->] (z); 
    \path (z) edge[ bend left=40,->] (x); 
    \node[]at (1.4*\W,\H){swap fee};
    \node[]at (1.4*\W,-\H){liquidity};
\end{tikzpicture}
\caption{Trader and liquidity provider (LP)}
\label{fig:relationDEX}
\end{subfigure}\hfill
\hfill
\begin{subfigure}[t]{0.46\linewidth}
\centering
\begin{tikzpicture}[state/.style={circle, draw, minimum size=1.2cm}]
    \node[state] at (0,0) (x) {\textbf{Trader}};%
    \node[state, right=of x, teal] at (\W,0) (z) {\textbf{MLP}}; %
    \node[state, right=of z, blue] at (5*\W,0) (y) {\textbf{Lender}}; %
    \path (x) edge[ bend left=40,->] (z); 
    \path (z) edge[ bend left=40,->] (x); 
    \path (y) edge[ bend left=30,->] (z); 
    \path (z) edge[ bend left=30,->] (y); 
    \node[]at (1.4*\W,\H){swap fee};
    \node[]at (1.4*\W,-\H){liquidity};
    \node[]at (4.55*\W,\H){position fee and collateral};
    \node[]at (4.7*\W,-\H){lend liquidity};
\end{tikzpicture}
\caption{Trader, margin liquidity provider (MLP), and lender or LP}
\label{fig:relationMLP}
\end{subfigure}
\caption{Interaction between market participants in an exchange.}
\label{fig:relation}
\end{figure*}
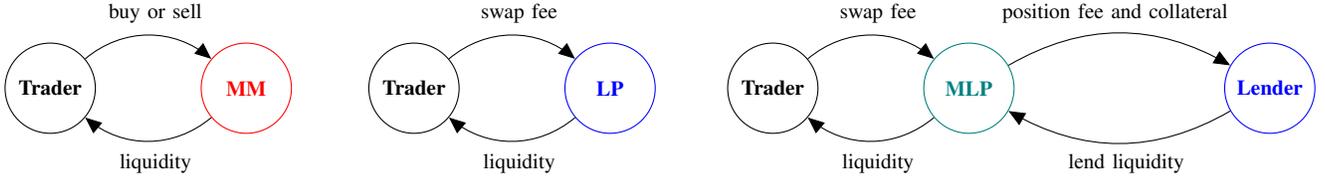

%% file: text/03preliminary.tex
Uniswap V2 \cite{uniswap} is the most popular DEX-AMM. LPs deposit two tokens in a Uniswap V2 pool and traders exchange a token for the other in the pool. To support this exchange mechanism, Uniswap V2 proposed CPMM (constant product market makers) where the product of the amount of two tokens is constant.

We denote the amount of each token deposited in the Uniswap V2 pool at time $t$ as $x_t, y_t$ for two exchangeable tokens $X$ and $Y$. Concretely, CPMM satisfies $x_ty_t = L^2$ for all time $t$, where $L$ is the conservative constant and denotes the liquidity of the pool. $L$ increases when LPs deposit their assets and decreases when LPs withdraw their assets. In Uniswap V2, swap fees distributed to LPs are reinvested, and $L$ increases. However, in this paper, we assume swap fees are collected individually as in the Uniswap V3 \cite{uniswapv3}.

\subsection{LP mechanism: add liquidity and remove liquidity}
\label{subsec:addremove}

LPs of the Uniswap V2 pool require a certain ratio of the amount of $X$ and $Y$ to deposit in the Uniswap V2 pool. We call this operation as \emph{adding liquidity}. When LPs withdraw, they receive $X$ and $Y$ in a certain ratio. We call this operation as \emph{removing liquidity}.

The ratio of the amount of $X$ and $Y$ LPs deposit or withdraw is the same as the ratio of the amount of $X$ and $Y$ deposited in the Uniswap V2 pool. Concretely, $\frac{\Delta x}{\Delta y}=\frac{x_t}{y_t}$ 
where $\Delta x$ and $\Delta y$ are the amount of $X$ and $Y$ an LP deposits or withdraws at time $t$, respectively.

The liquidity of the pool, $L$, increases when an LP deposits $\Delta x$ of $X$ and $\Delta y$ of $Y$. If the ratio of assets deposited by LPs to assets deposited in the pool is $d$, $L$ increases to $L(1+d)$. Concretely, 
\begin{align*}
    (x_t+\Delta x)(y_t+\Delta y) &= (x_t + dx_t)(y_t+d y_t)\\
    &=(1+d)^2x_ty_t\\
    &=\left((1+d)L\right)^2,
\end{align*}
where $d=\frac{\Delta x}{x_t}=\frac{\Delta y}{y_t}$.

\subsection{Exchange mechanism}

The exchange ratio is determined by the CPMM mechanism. When traders sell $\Delta x$ of $X$, $(1-\phi)\Delta x$ of $X$ are distributed to LPs in proportion to the amount LPs deposit and $\phi \Delta x$ of $X$ are added on the pool, where $\phi$ is set close to $1$. Then, $(x_t + \phi \Delta x)(y_t - \Delta y) = x_ty_t$ where $\Delta y$ is the amount of $Y$ extracted to the trader. Concretely,  
\begin{align*}
    \Delta y = \frac{y_t \phi \Delta x}{x_t + \phi \Delta x}.
\end{align*}
For simplicity, we set $\phi=1$ in this paper.

\subsection{Marginal price}

The price of the token in Binance, which is the largest CEX, can be globally measured by stablecoin, USDC. On the other hand, the price of each token in the Uniswap V2 pool is measured by the other token. In this regard, we define the exchange ratio of a very small amount of $Y$ to $X$ as the marginal price of $Y$ with respect to $X$. The marginal price of $Y$ with respect to $X$ can be represented with the amount of $X$ and $Y$ deposited in the Uniswap V2 pool as in \Cref{prop:marginalprice}. 

\input{prop/marginalprice.tex}
The amount of $X$, $x_t$ and the amount of $Y$, $y_t$ can be represented by the marginal price of $Y$ with respect to $X$, $p_t$ as follows: 
\begin{align}
    \label{eq:amountbypool}
    x_t = L\sqrt{p_t},
\quad    y_t = \frac{L}{\sqrt{p_t}}
\end{align} 

The ratio of marginal price between time $t$ and time $t'$ can be represented by the amount of $X$ at time $t$ and $t'$. Concretely, 
\begin{align}
    \frac{p_{t'}}{p_t} &=\frac{x_{t'}^2L^2}{L^2x_t^2}=\frac{{x_{t'}}^2}{{x_t}^2}.
    \label{eq:mpratio}
\end{align}

\subsection{Interpretation as an order book}

The trade amount in LOB is quoted at a certain price while the trade amount in DEX-AMM is quoted in the price range. To be more specific, the quoted amount of $Y$ between $[p_a, p_b]$ can be written as 
\begin{align*}
    \int_{p_a}^{p_b} \underbrace{\left( -\frac{dy}{dp} \right)}_{\text{quoted amount of $Y$ at  }p} dp = y_a - y_b
\end{align*}
In Uniswap V2, $y_a - y_b = L\left(\frac{1}{\sqrt{p_a}} - \frac{1}{\sqrt{p_b}}\right)$ by \Cref{eq:amountbypool}.
Note, the sum of quoted amount between price ranges is proportional to the liquidity, $L$.

\subsection{Divergence loss}
\label{divloss}
Divergence loss, also known as impermanent loss, is the opportunity cost for providing a pair of assets to the pool compared to simply holding them. The divergence loss is strictly determined by swap fees from traders to LPs and the difference between a deposited asset and a withdrawn asset. In this paper, however, we consider the profits from swap fees apart from the divergence loss for simplicity. The divergence loss is the value difference between a deposited asset and a withdrawn asset when LPs withdraw their assets.

%$$V(X_{init})=V(X_{final}) \text{ by definition of invariant}$$
%$$V(aX_{init}+(1-a)X_{final}) \leq aV(X_{init})+(1-%a)V(X_{final}) = V(X_{init})$$ by definition of convex
%therefore $$V(X_{init}+\Delta)\leq V(X_{init})$$
%For simplicity, let $$\frac{b}{c}<0, \Delta=(b, c), X_{init}=(x_{init}, y_{init})$$
%by Taylor's 1st expansion
%$$V(X_{init})+b\frac{\partial V} {\partial b}|_{init}+c\frac{\partial V} {\partial c}|_{init}\leq V(X_{init})$$
%$$-\frac{b}{c}p_{init}+1\leq 0$$
%By definition price is 
%$$p=-\frac{\frac{\partial V} {\partial b}}{\frac{\partial V}{\partial c}}$$
%\begin{align}
%    DL = (x_{init}+b)p|_{final}+y_{init}+c \\
%    -(x_{init}p|_{init}+y_{init})\\
%    = (x_{init}+b)p|_{final}-x_{init}p|_{init}+c
%\end{align}
%It is determined by factors including swap fee ratio, swap volume, change of marginal price, initial reserve amount, and final reserve amount. These factors are changed by swaps on a fixed AMM curve.

Suppose an LP deposits the asset at time $t_0$ and withdraws at time $t_1$. We denote the asset of the LP as a tuple of the amount of $X$ and $Y$. For example, we denote the deposited asset of the LP at time $t_0$ and the withdrawn asset of the LP at time $t_1$ as $(\Delta x_0, \Delta y_0)$ and $(\Delta x_1, \Delta y_1)$, respectively. 

The value of a withdrawn asset with respect to $X$ is $\Delta x_1+ p_1\Delta y_1$ where $p_1$ is the marginal price of $Y$ with respect to $X$ at time $t_1$.  On the other hand, suppose the LP holds the asset $(\Delta x_0, \Delta y_0)$ without deposit. Then, the value of the held asset with respect to $X$ at time $t_1$ is $\Delta x_0+ p_1 \Delta y_0$.

The difference between the value of a withdrawn asset with respect to $X$ at time $t_1$, $\Delta x_1+ p_1\Delta y_1$, and the value of the deposited asset with respect to $X$ at time $t_1$, $\Delta x_0+ p_1\Delta y_0$ is the divergence loss with respect to $X$ and we denote the divergence loss as $\delta$. $\delta$ is strictly negative as shown in \Cref{prop:divergence}.

\input{prop/divergence.tex}

LPs can still make a profit despite the divergence loss since the LPs are rewarded with swap fees paid by traders. Because divergence loss increases quadratically as marginal price deviates from the initial value, rational market participants will only provide liquidity in DEX only when they expect a sideways market. 
%Note that, LPs are similar to HFT (High-Frequency Trading) traders in CEX since the HFT strategy is applied when a sideways market is expected.

%% file: prop/marginalprice.tex
\begin{proposition}
\label{prop:marginalprice}
The marginal price of $Y$ with respect to $X$ at time $t$ is the ratio of the amount of $X$ and $Y$ deposited in the Uniswap V2 pool. As a corollary, the marginal price of $Y$ with respect to $X$ at time $t$ is the ratio of the amount of $X$ and $Y$ LPs deposit or withdraw in the Uniswap V2 pool at time $t$.
\end{proposition}
\begin{proof}
We first denote the marginal price of $Y$ with respect to $X$ at time $t$ as $p_t$. Then, $p_t=-\frac{\partial x_t}{\partial y_t}$ by the definition. Therefore, 
\begin{align*}
    p_{t}= -\frac{\partial x_t} {\partial y_t}= \frac{x_t}{y_t}=\frac{x_t^2}{L^2}=\frac{L^2}{y_t^2}.
\end{align*}

As noted in \Cref{subsec:addremove}, the ratio of the amount of $X$ and $Y$ an LP deposit or withdraw is the same as the ratio of the total amount of $X$ and $Y$ deposited in the Uniswap V2 pool. Concretely, $\frac{x_t}{y_t} = \frac{\Delta x_t} {\Delta y_t}$ where $\Delta x_t$ and $\Delta y_t$ are the amount of $X$ and $Y$ the LP deposit or withdraw at time $t$, respectively. Therefore, $p_t = \frac{\Delta x_t} {\Delta y_t}$.
\end{proof}

%% file: prop/divergence.tex
\begin{proposition}
\label{prop:divergence}
The divergence loss with respect to $X$ at time $t_1$, $\delta$ is $-\frac{(\Delta x_1 - \Delta x_0)^2}{\Delta x_0}$ where $\Delta x_1$ is the withdrawn $X$ at time $t_1$ and $\Delta x_0$ is the deposited $X$ at time $t_0$.

\end{proposition}
\begin{proof}
The divergence loss, $\delta$ is the difference between the value of a withdrawn asset with respect to $X$ at time $t_1$ and the value of the deposited asset with respect to $X$ at time $t_1$ or 
$\underbrace{\left(\Delta x_1 + p_1\Delta y_1 \right)}_{\text{withdrawn asset}} - \underbrace{\left(\Delta x_0 + p_1\Delta y_0\right)}_{ \text{deposited asset}}$. First, $p_1 = \frac{\Delta x_1} {\Delta y_1}$ by \Cref{prop:marginalprice}. Therefore,
\begin{align*}
    \delta  & = \underbrace{\left(\Delta x_1 + p_1\Delta y_1 \right)}_{\text{withdrawn asset}} - \underbrace{\left(\Delta x_0 + p_1\Delta y_0\right)}_{ \text{deposited asset}}\\
    & = \Delta x_1 - \Delta x_0 + p_1 \left(\Delta y_1 - \Delta y_0\right) \\
    & = \Delta x_1 - \Delta x_0 + \frac{\Delta x_1}{\Delta y_1} \left(\Delta y_1 - \Delta y_0\right) \\
    & = \Delta x_1 - \Delta x_0 + \Delta x_1 \left(1- \frac{\Delta y_0} {\Delta y_1}\right).
\end{align*}

The product of the amount of two tokens should be constant in CPMM. Therefore, $\Delta x_0 \Delta y_0 = \Delta x_1 \Delta y_1$ and $\frac{\Delta y_0}{\Delta y_1} = \frac{\Delta x_1} {\Delta x_0}$.

\begin{align*}
    \delta  &= \Delta x_1 - \Delta x_0 + \Delta x_1 \left(1- \frac{\Delta y_0} {\Delta y_1}\right)  \\
    & = \Delta x_1 - \Delta x_0 + \Delta x_1 \left(1- \frac{\Delta x_1} {\Delta x_0}\right) = - \frac{\left(\Delta x_1 - \Delta x_0\right)^2}{\Delta x_0}. 
\end{align*}
\end{proof}

%% file: text/04method.tex
We propose a novel position to reduce the risk of divergence loss for LPs in DEX-AMM. We first introduce a new concept called margin liquidity. We then propose a highly profitable position that attracts market participants by utilizing this margin liquidity. Finally, we introduce a modified version of the proposed position that is still highly profitable but at the same time reduces the risk of divergence loss. 
% provide more liquidity by lumping the risk of the divergence loss onto virtual margin liquidity providers.

\input{fig/liquidation}
\subsection{Margin liquidity}
\label{subsec:margin}

Can we provide more liquidity than the assets we have? Similar to margin trading, we define \emph{margin liquidity} as the liquidity introduced by borrowing assets from lenders with collateral. If the divergence loss of the lent liquidity exceeds the collateral, the lent liquidity and collateral are automatically withdrawn to lenders.

We can apply this concept of margin liquidity to any AMM where divergence loss exists \cite{uniswap, uniswapv3, curvev1, balancer}. In this paper, we instantiate margin liquidity on top of the Uniswap V2 framework. We suppose MLPs in Uniswap V2 provide $(l\Delta x_0, l\Delta y_0)$ when the collateral is $(\Delta x_0,\Delta y_0)$ and the leverage ratio is $l$. The ratio of $\Delta x_0$ and $\Delta y_0$ should be fixed as the current marginal price of $Y$ with respect to $X$ by \Cref{prop:marginalprice}.

\subsection{Margin liquidity-providing position}
\label{subsec:MLP}

We propose a \emph{margin liquidity-providing position} using the concept of margin liquidity. This position requires lenders to lend assets to market participants who open the position. Lenders can be the exchange or other market participants. When market participants open the margin liquidity-providing position, market participants borrow the assets from the lenders with collateral and provide liquidity by depositing the lent assets. 

We refer to the market participant who opens margin liquidity-providing positions as the margin liquidity provider (MLP). MLPs receive swap fees from traders in return for providing liquidity when a margin liquidity-providing position opens. The interactions between market participants (MLP, lender, and trader) are illustrated in \Cref{fig:relationMLP}.

The margin liquidity-providing position is liquidated when the divergence loss exceeds the collateral. The position can also be closed by the action of the MLP. When the margin liquidity-providing position closes, the deposited assets are withdrawn and returned to the lenders, along with a portion of the collateral corresponding to the divergence loss. Therefore, from the perspective of the lender, the value of the lent asset is equal to the value of the returned asset when the asset is returned. In detail, \Cref{prop:margin} states the ratio of the divergence loss to the collateral when the position closes. It also states the condition when the marginal liquidity-providing position is liquidated. Furthermore, \Cref{fig:liquidation} illustrates the rate of marginal price change where the marginal liquidity-providing position is not liquidated for each leverage ratio.

\input{prop/margin.tex}

Margin liquidity-providing position can be applied to both DEX and CEX. In DEX, a margin liquidity-providing position provides liquidity within a specific price range as shown in the liquidity curve of \Cref{fig:margin liquidity}. Liquidity in the price range is increased by the margin liquidity-providing position. The position yields a similar effect in CEX by facilitating market-making in the order book within the price range.

Note that margin liquidity-providing positions can be interpreted as betting on the sideways market while traditional margin trading bets on the bull market or bear market. Margin liquidity-providing positions have no potential danger of cascading liquidation as in long and short positions of margin trading since closing the margin liquidity-providing positions just reduces the liquidity in exchange without price change.
\begin{figure}[t]
\centering %
\begin{subfigure}[t]{0.5\linewidth}
\input{fig/MLP}
\caption{MLP}
\label{fig:margin liquidity}
\end{subfigure}\hfill
\begin{subfigure}[t]{0.5\linewidth}
\input{fig/VMLP}
\caption{VMLP }
\label{fig:virtual margin liquidity}
\end{subfigure}\hfill
\caption{Example of liquidity curves when a margin liquidity-providing position opens (left) and a virtual margin liquidity-providing position opens (right) on the Uniswap V2. Actual liquidity provided in the selected price range is increased by margin liquidity in \Cref{fig:margin liquidity}, while
actual liquidity provided in the selected price range is the same before and after in \Cref{fig:virtual margin liquidity}.
} 
\label{fig:mlpvmlp}
\end{figure}

\subsection{Virtual Margin liquidity-providing position}
\label{subsec:VMLP}

Although the introduction of the margin liquidity-providing position increases the total liquidity of the market by attracting risk-taking market participants, traditional LPs still have to suffer from divergence loss, which is the main source of the liquidity shortage in DEX-AMM. To address this issue, we propose a \emph{virtual} margin liquidity-providing position to lower the risk of divergence loss. Virtual margin liquidity-providing position borrows liquidity from LPs while margin liquidity-providing position borrows liquidity from lenders. 

While a margin liquidity-providing position increases the actual liquidity provided to the market, a virtual margin liquidity-providing position does not change the actual liquidity provided to the market. This is because the liquidity provided by the virtual liquidity-providing position comes from the liquidity previously provided by LPs in the same DEX pool. \Cref{fig:mlpvmlp} illustrates this difference between a margin liquidity-providing position and a virtual liquidity-providing position.

We name the market participants who open this position \emph{virtual margin liquidity providers} (VMLP). VMLPs take ownership of the liquidity provided by LPs and are rewarded with swap fees based on their ownership. However, when VMLPs close their positions, VMLPs compensate LPs for any divergence loss of the lent liquidity incurred during the loan period with their collateral. As a result, the virtual margin liquidity-providing positions are liquidated when the divergence loss of the lent liquidity exceeds the collateral.

LPs are rewarded by position fees from VMLPs proportional to the loan period, while the divergence loss of the lent liquidity is compensated. Therefore, their assets increase if the value of assets goes sideways. The interactions between the introduced market participants (VMLP, LP, and trader) are illustrated in \Cref{fig:relationMLP}.

Risk-averse market participants would prefer LP with low-risk and low returns. On the other hand, risk-taking market participants would prefer VMLP with high-risk and high returns.

%% file: fig/liquidation.tex
\begin{figure}
\centering
\begin{tikzpicture}
\begin{axis}[
axis on top=false,
width=0.9\linewidth,
height=0.6\linewidth,
xmin=0.3, xmax=4.2,
ymin=-4.1, ymax=4,
xtick={0.301, 1, 2, 3, 4},
xticklabels={2, 10, 100, 1000, 10000},
ytick={-4,-3,-2,-1,0,1,2,3,4},
yticklabels={1/16, 1/8,1/4,1/2,1,2,4,8,16},
xlabel={$l$},
ylabel={$\frac{p_1}{p_0}$},
ylabel style={rotate=-90},
grid=major,
/pgf/number format/fixed,
/pgf/number format/precision=3,
domain=0:5
]
    \addplot[name path=B, red,no marks,very thick] {
    2*log2((10^\x+sqrt(2*10^\x-1))/(10^\x-1))
    };
    \addlegendentry{
    $(\frac{l+ \sqrt{2l-1}}{l-1})^2$
    }
    \addplot[name path=A, blue,no marks,very thick] {
    2*log2((10^\x-sqrt(2*(10^\x)-1))/(10^\x-1))
    };
    \addlegendentry
    { 
    $(\frac{l- \sqrt{2l-1}}{l-1})^2$
    }
    \addplot [purple!30, fill opacity=0.3] fill between [of = A and B];
\end{axis}
 \end{tikzpicture}
 \caption{The range of the rate of marginal price change ($\frac{p_1}{p_0}$) where margin liquidity-providing position is not liquidated for each leverage ratio, $l$.}
 \label{fig:liquidation}
 \end{figure}

%% file: prop/margin.tex
\begin{proposition}
    \label{prop:margin}
    Suppose a margin liquidity-providing position opens when the marginal price is $p_0$. When the marginal price is $p_1$, the ratio of the divergence loss to the collateral is $l\left(\sqrt{\frac{p_1}{p_0}} - 1\right)^2 \cdot \frac{1}{1+\frac{p_1}{p_0}}$. As a corollary, the margin liquidity-providing position is liquidated if the rate of marginal price change, $\frac{p_1}{p_0}$ is bigger than $\left(\frac{l+ \sqrt{2l-1}}{l-1}\right)^2$ or less than $\left(\frac{l- \sqrt{2l-1}}{l-1}\right)^2$.
\end{proposition}
\begin{proof}

Suppose an MLP opens a margin liquidity-providing position when the marginal price is $p_0$ by depositing $(l\Delta x_0, l\Delta y_0)$ as liquidity when the collateral is $(\Delta x_0, \Delta y_0)$ and the leverage ratio is $l$. When the marginal price is $p_1$, the divergence loss is $\delta = \frac{ \left(l\Delta x_1-l\Delta x_0\right)^2}{l \Delta x_0}$, where $(l\Delta x_1, l \Delta y_1)$ is the deposited asset of the position. Note, $l\Delta x_0 \cdot l \Delta y_0 = l \Delta x_1 \cdot l\Delta y_1$. If the margin liquidity-providing position closes when the marginal price is $p_1$, the portion of the collateral corresponding to the divergence loss is returned to the lender. We denote the ratio of the divergence loss to the collateral as $r$. Concretely,
\begin{align*}
    r &= \frac{\delta}{\underbrace{\Delta x_0 + p_1 \Delta y_0}_{\text{collateral}}}\\
    &= \frac{ \left(l\Delta x_1 - l \Delta x_0\right)^2}{l\Delta x_0}\cdot \frac{1}{\Delta x_0 + p_1 \Delta y_0} \tag{by \Cref{prop:divergence}}\\
    &=l\frac{ \left(\Delta x_1 - \Delta x_0\right)^2}{\Delta x_0}\cdot \frac{1}{\Delta x_0 + p_1\frac{\Delta x_0}{p_0}} \tag{by \Cref{prop:marginalprice}}\\
    %&=l\frac{ \left(\Delta x_1 - \Delta x_0\right)^2}{\Delta x_0}\cdot \frac{1}{\Delta x_0 + \Delta x_0\frac{p_1}{p_0}}\\
    %&=l\left(\frac{\Delta x_1}{\Delta x_0} - 1\right)^2 \cdot\frac{1}{1+\frac{p_1}{p_0} }\\
    &=l\left(\sqrt{\frac{p_1}{p_0}} - 1\right)^2 \cdot \frac{1}{1+\frac{p_1}{p_0}}. \tag{by \Cref{eq:mpratio}}
\end{align*}

The position is not liquidated when the portion $r\leq 1$. Concretely,
\begin{align*}
    1 &\geq \underbrace{l\left(\sqrt{\frac{p_1}{p_0}} - 1\right)^2 \cdot \frac{1}{1+\frac{p_1}{p_0}}}_{\coloneqq r}\\
    %1+\frac{p_1}{p_0}&\geq l\left(\sqrt{\frac{p_1}{p_0}} - 1\right)^2\\
    %0&\geq (l-1)\left(\sqrt{\frac{p_1}{p_0}}\right)^2 -2l \sqrt{\frac{p_1}{p_0}} + (l-1) \\
    \therefore \left(\frac{l+ \sqrt{2l-1}}{l-1}\right)^2 &\geq \frac{p_1}{p_0} \geq \left(\frac{l- \sqrt{2l-1}}{l-1}\right)^2
\end{align*}

The margin liquidity-providing position is liquidated if the rate of marginal price change (=$\frac{p_1}{p_0}$) is bigger than $\left(\frac{l+ \sqrt{2l-1}}{l-1}\right)^2$ or less than $\left(\frac{l- \sqrt{2l-1}}{l-1}\right)^2$.
\end{proof}

%% file: fig/MLP.tex
\centering
\begin{tikzpicture}
\begin{axis}[
axis on top=false,
width=4.8cm,
height=4.8cm,
xmin=0, xmax=5,
ymin=0, ymax=5,
xtick={2,4},
xticklabels={$x_0$, $2x_0$},
ytick={2,4},
yticklabels={$y_0$, $2y_0$},
xlabel={$X$},
ylabel={$Y$},
grid=major,
label style={font=\scriptsize},
tick label style={font=\scriptsize},
legend style={anchor=north east, at={(0.97,0.95)}, font=\scriptsize},
/pgf/number format/fixed,
/pgf/number format/precision=3,
domain=0:5
]
    \addplot[domain=2.828:5, blue, no marks, thick]{2/\x};
    \addlegendentry{Total liquidity}
     \addplot[domain=0.5:2, red, no marks]{1/\x};
    \addlegendentry{MLP liquidity}
    \addplot[domain=0:5, blue, dashed, thick]{2/\x};
    \addlegendentry{LP liquidity}
    \addplot[domain=0:0.707, blue, no marks, thick]{2/\x};
    \addplot[domain=0.8585:3.4641, blue, no marks, thick]{3/\x};
    \addplot[black, dashed] {4*x};
    \addplot[black, dashed] {1/4*x};
    % Fill area between paths
\end{axis}
\end{tikzpicture}

%% file: fig/VMLP.tex
\centering
\begin{tikzpicture}
\begin{axis}[
axis on top=false,
width=4.8cm,
height=4.8cm,
xmin=0, xmax=5,
ymin=0, ymax=5,
xtick={2,4},
xticklabels={$x_0$, $2x_0$},
ytick={2,4},
yticklabels={$y_0$, $2y_0$},
xlabel={$X$},
grid=major,
ylabel style={rotate=-90},
label style={font=\scriptsize},
tick label style={font=\scriptsize},
legend style={anchor=north east, at={(1.1,0.95)}, font=\scriptsize},
/pgf/number format/fixed,
/pgf/number format/precision=3,
domain=0:5
]
    \addplot[domain=0:5, blue, no marks, thick]{2/\x};
    \addlegendentry{Total liquidity}
    \addplot[domain=0.35355:1.414, red, no marks]{1/2/\x};
    \addlegendentry{VMLP liquidity}
    \addplot[domain=2.828:5, blue, dashed, thick]{2/\x};
    \addlegendentry{LP liquidity}
    \addplot[domain=0:0.707, blue, no marks]{2/\x}; 
    \addplot[domain=0.6123:2.449, blue, dashed, thick]{3/2/\x};
    \addplot[black, dashed, thick] {4*x};
    \addplot[black, dashed, thick] {1/4*x};
    % Fill area between paths
\end{axis}
\end{tikzpicture}

%% file: text/05relatedworks.tex
\subsection{Uniswap V3}
Uniswap V2 provides liquidity uniformly across the whole price range $(0, \infty)$. Hence, most of the liquidity is provided in the price range far from the current price. To deal with this problem, Uniswap V3 proposes concentrated liquidity which allows LPs to provide liquidity only on the bounded price range close to the current price. Therefore, the LPs in Uniswap V3 can provide the same liquidity to the market with a smaller capital.

Margin liquidity provides liquidity on a bounded price range like Uniswap V3. However, margin liquidity provides much more liquidity in the same price range than the 
concentrated liquidity in Uniswap V3 as shown in \Cref{subsec: capeff}.

%Uniswap V3 introduced a tick-like order book enabling LPs to set a price range in which their liquidities are activated. 

%There are trade-offs between Uniswap V2 and Uniswap V3. To make Uniswap V3 work, LPs  uninformed LPs don't rebalance their assets   Uniswap V2 LPs can make more profits 
%If, one of a pair of tokens in the Uniswap V3 pool can be bankrupt Unlike Uniswap V2

\subsection{dYdX}

% 1.dydx 는 dex에서 long short position 잡는 거 도와주는 프로토콜 we call margin trader open and close =
% 2.position 잡는 사람의 담보금을 바탕으로 dydx protocol 이 lender의 돈으로 대신 거래 (우리는 3rd party의존 안함) 포지션 종료시 lender한테는 interest fee => 우리도 position fee를 준다. lender 토큰 하나를 빌리는데 우리는 토큰 2개의 pair를 빌리고 divergence loss로 부터 방어 토큰 수량만 그대로돌려줌 
% 3. 포지션이 강제종료에 가까워질 때 담보금 넣지 않으면 liquidator에 의해 강제 청산 당할수 있음( automatic liquidation)

dYdX DEX \cite{dydx} provides perpetual margin tradings in DeFi. We call market participants who open the position for margin trading as margin traders. Margin traders buy or sell assets by leveraging their loans. On the other hand, MLPs leverage their loan to provide liquidity for other traders.

dYdX protocol buys or sells tokens using third-party DEX with the assets of lenders when margin traders open a long or short position with their collaterals. On the other hand, our margin liquidity-providing position does not require any third-party DEX to buy or sell tokens. 

%Margin traders open Its main players consist of lenders, margin traders, and liquidators in terms of their behaviors.
%Margin traders generally open a new position with the info of the loan and the buy order of a specified token with an amount. They can also close in any portion of the position at any price only if the counterpart positions exist enough to take a sell order to pay loans and interest fees to lenders.
%Lenders offer loans specified in the amount, the sort of tokens, and interest rates for margin traders to enable leveraging. dYdX protocol buys or sells tokens using 3rd party DEX with the assets of lenders in the capability of traders' collaterals to

\subsection{Aave V3}
Aave provides a pool-based crypto lending service in DeFi. In the Aave lending pool, lenders deposit crypto assets, and users can borrow the assets with their collaterals when Loan-to-Value (LTV) is below the limit set by the Aave protocol\cite{soklend, sokagg, aavev3, aaveproto2, aaveproto}. Unlike margin tradings or margin liquidity, the value of the collaterals is usually higher than the value of borrowed assets\cite{stablecoin2}. 

Aave requires a third party to liquidate the assets of users when LTV approaches the limit\cite{liquidation}. On the other hand, the position is liquidated without any support from a third party in the margin liquidity-providing position.

%The second one is a new concept named flashloan. Flashloan is designed for borrowing crypto assets without any collateral, only if the payback amounts of borrowed assets are guaranteed to pay loans with stable rates of interest and fees in one transaction. Aave protocol first suggested the flashloan by using the concept of atomic transaction in Ethereum Virtual Machine(EVM). 

%% file: text/06application.tex
\begin{figure*}[t]
\def\W{0.1}     %width
\def\H{0.1}     %height
\centering %
\begin{subfigure}[t]{0.23\linewidth}
\includegraphics[width=\linewidth,page=1]{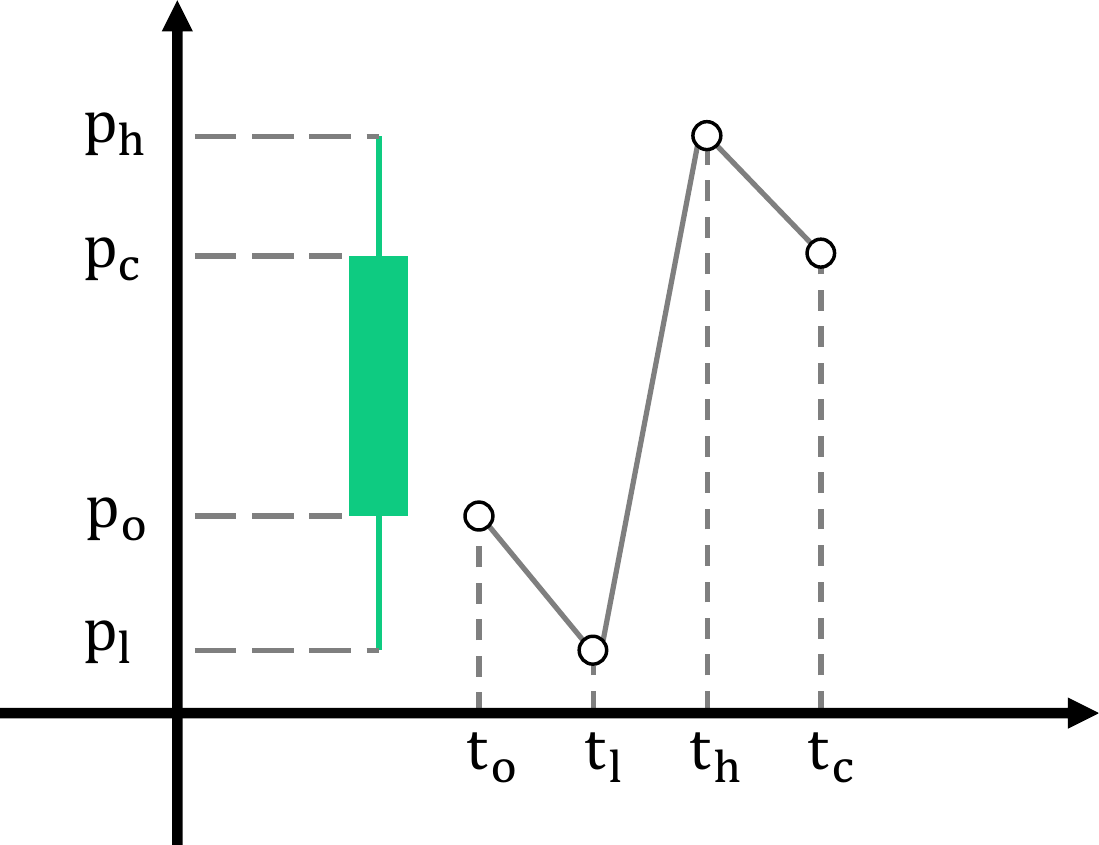}
\caption{$p_o<p_c$\\ 
$t_o<t_l<t_h<t_c$}
\label{fig:ohlcv1}
\end{subfigure}\hfill
\begin{subfigure}[t]{0.23\linewidth}
\includegraphics[width=\linewidth,page=2]{fig/ohlcv.pdf}
\caption{$p_o<p_c$\\ 
$t_o<t_h<t_l<t_c$}
\label{fig:ohlcv2}
\end{subfigure}\hfill
\begin{subfigure}[t]{0.23\linewidth}
\includegraphics[width=\linewidth,page=3]{fig/ohlcv.pdf}
\caption{$p_o>p_c$\\ 
$t_o<t_l<t_h<t_c$}
\label{fig:ohlcv3}
\end{subfigure}\hfill
\begin{subfigure}[t]{0.23\linewidth}
\includegraphics[width=\linewidth,page=4]{fig/ohlcv.pdf}
\caption{$p_o>p_c$\\ 
$t_o<t_h<t_l<t_c$}
\label{fig:ohlcv4}
\end{subfigure}
\caption{Four possible scenarios of the price candle.
} 
\label{fig:ohlcv}
\end{figure*}

We propose a representative strategy for MLPs who open and close the margin liquidity-providing position to maximize their profit. We backtest\cite{backtest} an MLP that opens and closes a margin liquidity-providing position to provide liquidity on the pool of $X$ and $Y$.

\subsection{Price}
We denote the USD price of $X$ and $Y$ at time $t$ as $p_x(t)$ and $p_y(t)$, respectively. For the backtesting, we bring the historical data of  $p_x(t)$, $p_y(t)$, and $p(t)$ from Binance. In detail, there are open price, close price, high price, and low price for each timestamp $t$ in OHLCV data. $p_x(t)$, $p_y(t)$, $p(t)$ are set as the open price in this paper.

We evaluate the asset of an MLP at time $t$ in USD terms. Then, we denote the value of the MLP's asset at time $t$ as $c(t)$ USD. Concretely, if the MLP has $\Delta x$ of $X$ and $\Delta y$ of $Y$, then $c(t) := p_x (t) \Delta x + p_y(t) \Delta y$. 

\subsection{Opening criteria}

The margin liquidity-providing position yields the maximum profit in a sideways market. Therefore, we define the price trend slope at time $t$ as $\alpha(t)$ in \Cref{def:pricetrend} and determine whether the market is in the uptrend, downtrend, or sideways market at each time $t$ with this measure. Note that $a(t)\gg 0$ indicates the market is in an uptrend, while $a(t)\ll 0$ indicates the market is in a downtrend. In this regard, we assume the MLP judges the market is in the sideways market when $|a(t)| \leq \alpha$ for a fixed hyperparameter $\alpha$ and opens the margin liquidity-providing position. 

\input{prop/pricetrendslope.tex}

We assume the MLP opens margin liquidity-providing position at time $t_o$ with the whole asset $c(t_o)$. Then, the MLP buys $\frac{p(t_o)c(t_o)}{p_y(t_o) + p_x(t_o) p(t_o)}$ of $X$ and $\frac{c(t_o)}{p_y(t_o) +p_x(t_o)p(t_o)}$ of $Y$ for the collateral at the price $p_x(t_o)$ and $p_y(t_o)$, respectively according to \Cref{prop:mlpbuy}. For simplicity, we suppose MLP buys  $\Delta x$ of $X$ and $\Delta y$ of $Y$. Then, MLP deposits $(l\Delta x, l\Delta y)$ with the collateral $(\Delta x, \Delta y)$. 

\begin{proposition}
\label{prop:mlpbuy}
When the asset of MLP is $c(t)$ USD at time $t$, MLP can buy $\frac{p(t)c(t)}{p_y(t) + p_x(t) p(t)}$ of $X$ and $\frac{c(t)}{p_y(t) +p_x(t)p(t)}$ of $Y$ as collateral where the marginal price of $Y$ with respect to $X$ is $p(t)$, and the marginal price of $X$ and $Y$ are $p_x(t)$ and $p_y(t)$, respectively.
\end{proposition}
\begin{proof}
When the asset of the MLP is $c(t)$, suppose the MLP buys $\Delta x$ of $X$ and $\Delta y$ of $Y$. Then, $c(t) = p_x(t) \Delta x  + p_y(t) \Delta y$.
Also, $p(t) = \frac{\Delta x}{\Delta y}$ by \Cref{prop:marginalprice}. Therefore, 
\begin{align*}
    \Delta x= \frac{p(t)c(t)}{p_y(t) + p_x(t) p(t)}, \quad
    \Delta y= \frac{c(t)}{p_y(t) +p_x(t)p(t)}
\end{align*}
\end{proof}

\subsection{PNL analysis}
\label{subsecsec:pnl}

We can estimate the profit and loss (PNL) for the MLP position by 1) estimating the swap fees from trading volumes on the provided liquidity by deposited assets $(l\Delta x, l\Delta y)$ and 2) computing the divergence loss due to the change of marginal price.

\subsubsection{Swap fees}

Swap fees returned to the MLP are proportional to the trade volume on the liquidity provided by the MLP. We denote the liquidity provided by the MLP as $L$, which is computed as $\sqrt{l \Delta x \cdot l\Delta y}$.

We track the trade volume on the liquidity provided by the MLP since the margin liquidity-providing position opens at time $t_o$. We denote the trade volume from both $x$ to $y$ and $y$ to $x$ since $t_o$ as $w_x(t)$ and $w_y(t)$, respectively. Then, $w_x(t) = \sum_{t'=t_o}^{t} v_x(t')$  where $v_x(t)$ is the trade volume from $X$ to $Y$ at time $t$ on the liquidity provided by the MLP. Similarly, $w_y(t) = \sum_{t'=t_o}^{t} v_y(t')$  where $v_y(t)$ is the trade volume from $Y$ to $X$ at time $t$ on the liquidity provided by the MLP. \Cref{prop:tradevolume} states the lower bound $v_x(t)$ and $v_y(t)$ for each time $t$. In the backtesting, we suppose the $v_x(t)$ and $v_y(t)$ are equal to the lower bound in \Cref{prop:tradevolume}. Then, we can compute the profit from the swap fees. The MLP receive $(1-\phi) w_x(t)$ of $X$ and $(1-\phi) w_y(t)$ of $Y$ until time $t$ as the swap fees. These fees are worth $(1-\phi) w_x (t) p_x(t) + (1-\phi) w_y (t) p_y(t)$ USD.
\input{prop/tradevolume.tex}

\subsubsection{Divergence loss} 

Suppose the MLP closes the margin liquidity-providing position at time $t$. The portion of collateral corresponding to the divergence loss is returned to the lender. We evaluate the value of the collateral left in the margin liquidity-providing position at time $t$ as the value of the withdrawn asset to the MLP if the position closes at time $t$. We denote the ratio of deposited collateral corresponding to the divergence loss at time $t$ as $r(t)$. Concretely, 
\begin{align*}
    r(t) = l\left(\sqrt{\frac{p(t)}{p(t_o)}} - 1\right)^2 \cdot \frac{1}{1+\frac{p(t)}{p(t_o)}}
\end{align*}
by \Cref{prop:margin}. Then, the withdrawn asset is $\left((1-r(t))\Delta x, (1-r(t))\Delta y\right)$ if the margin liquidity position closes at time $t$. Therefore, the value of the collateral left in the margin liquidity-providing position at time $t$ is $\left(1- r(t)\right) \left( p_x(t) \Delta x + p_y(t) \Delta y\right)$.

\subsubsection{PNL}

The estimated value of the MLP's asset at time $t$ is the sum of the value of collateral left in the margin liquidity position and the value of swap fees from traders. Concretely, 
\begin{align*}
    c(t) & = \underbrace{\left(1- r(t)\right) \left( p_x(t) \Delta x(t_o) + p_y(t) \Delta y(t_o)\right)}_{\text{the value of collateral left in the position}} \\
                  & + \underbrace{(1-\phi) w_x (t) p_x(t) + (1-\phi) w_y (t) p_y(t)}
                  _{\text{swap fees from traders}}.
\end{align*}
PNL at time $t$ is the difference between the estimated value of the MLP's asset at time $t$ and $t_o$. We denote the PNL at time $t$ as $\mathrm{PNL}(t)$. Then, $\mathrm{PNL}(t) = c(t) - c(t_o)$.

\subsection{Closing criteria}
There are three closing conditions for the margin liquidity-providing position. If one of the conditions is satisfied, the margin liquidity-providing position is voluntarily closed or liquidated. Firstly, if the absolute value of the price trend slope is larger than a fixed hyperparameter $\beta$ which is larger than $\alpha$, MLP judges the market converts from a sideways market to a trend market and closes the position. Secondly, if the PNL exceeds the stop loss, the MLP voluntarily closes the position. We denote the ratio of stop loss to the initial asset as $\gamma$. Then, the MLP closes the position when $\mathrm{PNL}(t) < - \gamma c(t_o)$. Finally, if the divergence loss exceeds collateral or $r(t)=1$, the position is liquidated.

%% file: prop/pricetrendslope.tex
\begin{definition}[Price trend slope]
\label{def:pricetrend}
We define the price trend slope at time $t$ as the slope of the linear trend line for the price data of the last $W$ time. Concretely, the price trend slope at time $t$, $a(t)$, is defined as
\begin{align*}
    a(t), b(t) = \arg\min_{a,b} \int_{t-W}^{t} \left(p(\tau) - a\tau - b\right)^2 d \tau.
\end{align*}
\end{definition}

%% file: prop/tradevolume.tex
\begin{proposition} 
\label{prop:tradevolume}
The lower bound of trade volume from $x$ to $y$ at time $t$ on the liquidity provided by MLP ($=\!L$) can be written as below:
\footnotesize
\begin{align}
\label{eq:lb1}
v_x(t) \geq \begin{cases} 
L\left(\sqrt{p_h(t)}  - \sqrt{p_l(t)}  \right)& \text{if $p_o<p_c$}\\
L\left(\sqrt{p_h(t)}  - \sqrt{p_o(t)}  \right) +
L\left(\sqrt{p_c(t)}  - \sqrt{p_l(t)}  \right)  
& \text{otherwise}
\end{cases}
\end{align}
\normalsize
Similarly, the lower bound of the trade volume from $y$ to $x$ at time $t$ on the liquidity provided by MLP ($=\!L$) can be written as below:
\footnotesize
\begin{align}
\label{eq:lb2}
v_y(t) \geq \begin{cases} 
L\left(\frac{1}{\sqrt{p_l(t)}}  - \frac{1}{\sqrt{p_o(t)}}  \right)  +
L\left(\frac{1}{\sqrt{p_c(t)}}  - \frac{1}{\sqrt{p_h(t)}}  \right) & \text{if $p_o<p_c$}\\
L\left(\frac{1}{\sqrt{p_l(t)}}  - \frac{1}{\sqrt{p_h(t)}}  \right)& \text{otherwise}
\end{cases}
\end{align}
\normalsize
\end{proposition}
\begin{proof}
We first denote the reference time for the open price $p_o$, close price $p_c$, high price $p_h$, and low price $p_l$ as $t_o$, $t_c$, $t_h$, and $t_l$, respectively. By the definition of OHLCV, $t_o< t_h, t_l <t_c$. Therefore, we can divide OHLCV into four possible cases according to two criteria: 1) $p_o<p_c$ or else and 2) $t_l<t_h$ or else as in \Cref{fig:ohlcv}.

The marginal price increase means traders swap $X$ to $Y$. On the contrary, the marginal price decrease means traders swap $Y$ to $X$. Note that when the price changes, the trade volume is at least the amount changed in the pool which can be represented by the marginal price by \Cref{eq:amountbypool}. 

\begin{itemize}
    \item In \Cref{fig:ohlcv1} and \Cref{fig:ohlcv3},
    \footnotesize
    \begin{align}
    \label{eq:OHLCV13}
    v_x(t) &\geq \underbrace{L\left(\sqrt{p_h(t)}  - \sqrt{p_l(t)}  \right)}_{t_l \to t_h}\\
    v_y(t) &\geq \underbrace{L\left(\frac{1}{\sqrt{p_l(t)}}  - \frac{1}{\sqrt{p_o(t)}}  \right)}_{ 
    t_o \to t_l
    }  
    + 
    \underbrace{L\left(\frac{1}{\sqrt{p_c(t)}}  - \frac{1}{\sqrt{p_h(t)}}  \right)}_{
    t_h \to t_c
    }\nonumber
    \end{align} 
    \normalsize
    \item In \Cref{fig:ohlcv2} and \Cref{fig:ohlcv4},
        \footnotesize
    \begin{align}
    \label{eq:OHLCV24}
    v_x(t) &\geq \underbrace{L\left(\sqrt{p_h(t)}  - \sqrt{p_o(t)}  \right)}_{t_o \to t_h}  +
    \underbrace{L\left(\sqrt{p_c(t)}  - \sqrt{p_l(t)}  \right)}_ {t_l \to t_c}\\
v_y(t) &\geq \underbrace{L\left(\frac{1}{\sqrt{p_l(t)}}  - \frac{1}{\sqrt{p_h(t)}}  \right)}_{t_h \to t_l}\nonumber
    \end{align}
     \normalsize
\end{itemize}

\Cref{fig:ohlcv1} and \Cref{fig:ohlcv2} are indistinguishable from OHLCV data. Therefore, we estimate $v_x(t)$ and $v_y(t)$ by the lower bound of \Cref{eq:OHLCV13} and \Cref{eq:OHLCV24}.
\end{proof}

%% file: table/result.tex
%% v1
% \begin{table}[h!]
%     \begin{adjustbox}{max width=\columnwidth}
%     \centering
%     \begin{tabular}{c l c c c}
%         \toprule
%         period     & Method                  & Sharpe ratio $\uparrow$ & MDD $\downarrow$  & ROR$\uparrow$\\
%         \midrule
%         20Q1       & Baseline                & -0.21        & 0.666 & 0.95    \\
%         \cmidrule(r){2-5}
%                    &Ours ($l\!=\!1$)         & \textbf{2.89}         & \textbf{0.311} & \textbf{1.56}\\
%                    & Ours ($l\!=\!10$)       & 2.74         & 0.328 & 1.47\\
%                    & Ours ($l\!=\!100$)      & -0.80        & 0.437 & 0.91\\
%         \midrule
%         21Q1    & Baseline &   3.72       & 0.313 & 2.31\\
%         \cmidrule(r){2-5}
%                    & Ours ($l\!=\!1$)     &  4.21      & \textbf{0.301} & \textbf{2.14}\\
%                    & Ours ($l\!=\!10$)     &  \textbf{4.28}      & 0.302 & 2.00\\
%                    & Ours ($l\!=\!100$)     &  -1.16      & 0.465 & 0.86\\
%         \bottomrule
%     \end{tabular}
%     \caption{Backtesting results of our representative strategy and  in several periods (20Q1, 21Q1). We set $\alpha=0.1, \beta=0.2, \gamma=0.01$ and swap fee is $0.1\%$. We set time window $W=2 \mathrm{h}$ to calculate price trend slope defined as \Cref{def:pricetrend} 
%     }
%     \label{tab:simul}
%     \end{adjustbox}
% \end{table}

%% v2
\begin{table}[h!]
    \begin{adjustbox}{max width=\columnwidth}
    \centering
    \begin{tabular}{c l c c c}
        \toprule
        period  & Method & Sharpe ratio $\uparrow$ & MDD $\downarrow$  & ROR$\uparrow$\\
        \midrule
        20Q1    & Baseline & -0.21 & 0.67 & 0.95 \\ %p1
        \cmidrule(r){2-5}
                & Ours ($l\!=\!3$) & 3.31 & \textbf{0.32} & 1.84\\
                & Ours ($l\!=\!10$) & 3.32 & \textbf{0.32} & 1.84\\
                & Ours ($l\!=\!100$) & \textbf{3.53} & 0.33 & 1.91\\
                & Ours ($l\!=\!1000$) & 1.80 & 0.79 & \textbf{2.07}\\
        \midrule
        20Q3    & Baseline & 1.82 & 0.29 & 1.39\\ %p3
        \cmidrule(r){2-5}
                & Ours ($l\!=\!3$) & 2.20 & \textbf{0.21} & 1.84\\
                & Ours ($l\!=\!10$) & 2.22 & \textbf{0.21} & 1.84\\
                & Ours ($l\!=\!100$) & \textbf{2.35} & 0.22 & 1.91\\
                & Ours ($l\!=\!1000$) & -2.11 & 0.89 & \textbf{2.07}\\
        \midrule
        21Q1    & Baseline & 3.72 & 0.31 & 2.31\\ %p5
        \cmidrule(r){2-5}
                & Ours ($l\!=\!3$) & 4.04 & \textbf{0.30} & \textbf{2.55}\\
                & Ours ($l\!=\!10$) & 4.06 & \textbf{0.30} & 2.52\\
                & Ours ($l\!=\!100$) & \textbf{4.11} & 0.32 & 2.28\\
                & Ours ($l\!=\!1000$) & -3.42 & 0.92 & 0.42\\
        \bottomrule
    \end{tabular}
    \caption{Backtesting results of our representative strategy and  in several periods (20Q1, 20Q3, 21Q1). We set $\alpha=0.1, \beta=0.2, \gamma=0.05$ and swap fee is $0.15\%$. We set time window $W=2 \mathrm{h}$ to calculate price trend slope defined as \Cref{def:pricetrend} 
    }
    \label{tab:simul}
    \end{adjustbox}
\end{table}

%% file: text/07experiments.tex
\subsection{Backtesting for representative strategy}

We proposed our representative strategy in \Cref{sec:app}. In this section, we backtest our representative strategy with historical data from several periods (20Q1, 20Q3, 21Q1). We consider the margin liquidity-providing strategy on the ETH/BTC market and utilize `5m' Binance OHLCV data for ETH/USDC, BTC/USDC, and ETH/BTC.

We evaluate our representative strategy with Sharpe ratio, MDD, and ROR. Note that we suppose the risk-free rate in the Sharpe ratio is computed for the case where the USD is deposited in the bank with an annual interest rate of $8\%$. For the baseline, we propose an ETH and BTC holding position as baseline, where the ratio of the holding amount of ETH and BTC is the marginal price of ETH with respect to BTC following \Cref{prop:marginalprice} \cite{stylized}. \Cref{tab:simul} shows that our strategy can outperform the baseline with a higher Sharpe ratio, a smaller MDD, and a higher ROR if leverage ratio is set appropriately. For example, in 20Q1, the Sharpe ratio, MDD, and ROR for our strategy when $l=100$ are $3.53$, $0.33$, and $1.91$, while the baseline achieves $-0.21$, $0.67$, and $0.95$.

\subsection{Capital efficiency comparison}
\label{subsec: capeff}

In this paper, we use the term capital efficiency to denote how much liquidity can be provided with the same amount of assets. We aim to compare the capital efficiency of our proposed margin liquidity with the concentrated liquidity in Uniswap V3. Note that our proposed margin liquidity and concentrated liquidity are both extensions of Uniswap V2.

\subsubsection{Capital efficiency of concentrated liquidity}
We first suppose liquidity is provided in the price range $[p_l, p_u]$ in the Uniswap V3. Then, 
\begin{align*}
    (x+ L\sqrt{p_l})(y+\frac{L}{\sqrt{ p_u}})=L^2 
\end{align*}
where $x$ and $y$ are the number of $X$ tokens and $Y$ tokens, respectively. Also, the marginal price $p$ of $Y$ token with $X$ token can be computed as in \Cref{eq:uniswapv3price}.
\begin{align}
    \label{eq:uniswapv3price}
    p&= -\frac{dx}{dy} = \frac{L^2}{\left(y+\frac{L}{\sqrt{ p_u}}\right)^2} = \frac{ \left(x+L\sqrt{ p_l}\right)^2 }{L^2}
\end{align}
Then, $x=0$, $y=L\left(\frac{1}{\sqrt{p_l}} -\frac{1}{\sqrt{p_u}}\right)$ when the marginal price is $p_l$ and $x= L \left(\sqrt{p_u} -\sqrt{p_l}\right)$, $y=0$ when the marginal price is $p_u$. Therefore, Uniswap V3 provides $L \left(\sqrt{p_l} -\sqrt{p_u}\right)$ of $X$ token and $L\left(\frac{1}{\sqrt{p_l}} -\frac{1}{\sqrt{p_u}}\right)$ of $Y$ token in $[p_l,p_u]$.

Suppose the deposited assets in the Uniswap V3 pool are withdrawn at the marginal price $\sqrt{p_lp_u}$ and are redeposited to the Uniswap V2 pool. For brevity, let's denote the $\sqrt{p_lp_u}$ as $p_b$. Then, the assets amount of $\left(L \left(\sqrt{p_b} -\sqrt{p_l}\right), L \left(\frac{1}{\sqrt{p_b}} - \frac{1}{\sqrt{p_u}}\right)\right)$ are withdrawn and redeposited to the Uniswap V2 pool. The liquidity of the Uniswap V2 pool becomes \[ 
\sqrt{  
L \left(\sqrt{p_b} -\sqrt{p_l}\right)\cdot L \left(\frac{1}{\sqrt{p_b}} - \frac{1}{\sqrt{p_u}}\right)
}.
\]
We denote the liquidity of the Uniswap V2 pool as $L_0$. Similarly in the Uniswap V3, $x = L_0\sqrt{p_l}$, $y=\frac{L_0}{\sqrt{p_l}}$ when the marginal price is $p_l$ and  $x = L_0\sqrt{p_u}$, $y=\frac{L_0}{\sqrt{p_u}}$ when the marginal price is $p_u$ by \Cref{eq:amountbypool}. Therefore, Uniswap V2 provides $L_0 \left(\sqrt{p_l} -\sqrt{p_u}\right)$ of $X$ token and $L_0\left(\frac{1}{\sqrt{p_l}} -\frac{1}{\sqrt{p_u}}\right)$ of $Y$ token in the price range $[p_l,p_u]$.

The ratio of $L_0$ and $L$ represent the ratio of the number of tokens provided by the Uniswap V2 pool and Uniswap V3 pool in the price range $[p_l, p_u]$: 
\begin{align*}
    \frac{L_0}{L}= \sqrt{\left(\sqrt{p_b} -\sqrt{p_l}\right)\left(\frac{1}{\sqrt{p_b}} - \frac{1}{\sqrt{p_u}}  \right) } 
    %&= \sqrt{\left(1 -\sqrt{\frac{p_l}{p_b}}\right)\left(1 - \sqrt{ \frac{p_b}{p_u}}  \right) }   \\
    = 1 -\left(\frac{p_l}{p_u}\right)^{\frac{1}{4}}.
\end{align*}
The ratio above implies that concentrated liquidity in Uniswap V3 is $\frac{1}{1-\left(\frac{p_l}{p_u}\right)^{\frac{1}{4}}} \times$ more capital efficient than Uniswap V2.

\subsubsection{Capital efficiency of Margin liquidity}

Suppose we open a margin liquidity-providing position instead of re-depositing the asset to the Uniswap V2 pool. We set the leverage ratio of margin liquidity-providing position as $l$ such that the position remains open without liquidation in the price range $[p_l, p_u]$. This in turn means that $\left(\frac{l+ \sqrt{2l-1}}{l-1}\right)^2 \geq \frac{p_u}{p_b}$ and $\left(\frac{l- \sqrt{2l-1}}{l-1}\right)^2 \leq \frac{p_l}{p_b}$ by \Cref{prop:margin}. Then, the maximum leverage ratio possible is 
$\frac{ \left(\frac{p_u}{p_l}\right)^{\frac{1}{2}} + 1}{\left( \left(\frac{p_u}{p_l}\right)^{\frac{1}{4}}-1 \right)^2}$.

Margin liquidity is $l\times$ more capital efficient than Uniswap V2 since $l$ times more assets are deposited in the margin liquidity. In the best case, margin liquidity is $\frac{ \left(\frac{p_u}{p_l}\right)^{\frac{1}{2}} + 1}{\left( \left(\frac{p_u}{p_l}\right)^{\frac{1}{4}}-1 \right)^2} \times$ more capital efficient than Uniswap V2.

\input{fig/capitaleff.tex}
\cref{fig:captialeff} shows the capital efficiency of Uniswap V3 and margin liquidity using the results above. In particular, When $\frac{p_u}{p_l}=1.001$, concentrated liquidity in Uniswap V3 is $4K$ times more efficient than Uniswap V2 while margin liquidity is $32M$ times more efficient than Uniswap V2.

% crash 사건들 (maker dao, ethereum dao, luna) 우리 DEX에서 했었으면 어떻게 달라졌을까 우리는 trading volume으로 계산 3개에 대한 충격테스트, 일반적인 AMM UniswapV2만 사용했을때, V3, curve, dodo 정의한 메트릭들 이랑 비교했을때 우리것이 더 우수하다. 우리는 판다는 개념이 아리나 유동성을 뺀다는 개념 

% Liqudity 정의: AMM은 그대로 CEX:  long: 21초, short 22/5/7 ~ 5/11 횡보장: 5월 말 
% 20/6/12 ~ 20/7/20 횡보
% 2018/01/ ~ 12 short 
%
% Low peak 2021-07-21, 2020-03-16, 2021-010
% 2017-0
%
% supertrend
% period 1 
% 2021-02-23 - 2021-05-13 sideway 
% 2021-07-21 low peak
%
% period2 
% 2019-05-10- 2020-02-26 sideway
% 2020-03-16 low peak
%
% period3
% 2020-05-01  2020-10-20 sideway
% 2021-01-01  
% period4 
% 2017-01-04 2017-05-03 sideway
% 2017-06-12

%% file: fig/capitaleff.tex
\begin{figure}
\centering
\begin{tikzpicture}
\begin{axis}[
axis on top=false,
width=0.95\linewidth,
height=0.7\linewidth,
xmin=-3.5, xmax=3,
ymin=-1, ymax=8,
xtick={-3, -2, -1, 0, 0.9542, 1.9956, 2.9996},
xticklabels={1.001, 1.01, 1.1, 2, 10, 100, 1000},
ytick={0,1,2,3,4,5,6,7},
yticklabels={1,10,100,1K,10K,100K,1M, 10M},
xlabel={$\frac{p_u}{p_l}$},
ylabel={Capital efficiency},
ylabel style={rotate=0},
grid=major,
legend style={anchor=north east, at={(0.95,0.95)}},
/pgf/number format/fixed,
/pgf/number format/precision=3,
domain=-3.5:3
]
    \addplot[name path=B,blue,no marks,very thick] {
    log10((sqrt((1+10^\x))+1)/((sqrt(sqrt((1+10^\x)))-1)^2)
    };
    
    \addlegendentry{Margin liquidity}
    \addplot[name path=A, red,no marks,very thick] {
    log10( (sqrt(sqrt((1+10^\x))))/( sqrt(sqrt((1+10^\x)))-1 ))
    };
    \addlegendentry{Concentrated liquidity}
    %\addplot[scatter,only marks, black
    %scatter src=explicit symbolic]%
    %table[meta=label] {
    %x y label
    %-3.602 3.602 
    %};
\end{axis}
 \end{tikzpicture}
 \caption{Comparison of capital efficiency over Uniswap V2 between concentrated liquidity in Uniswap V3 and our proposed margin liquidity}
 \label{fig:captialeff}
\end{figure}
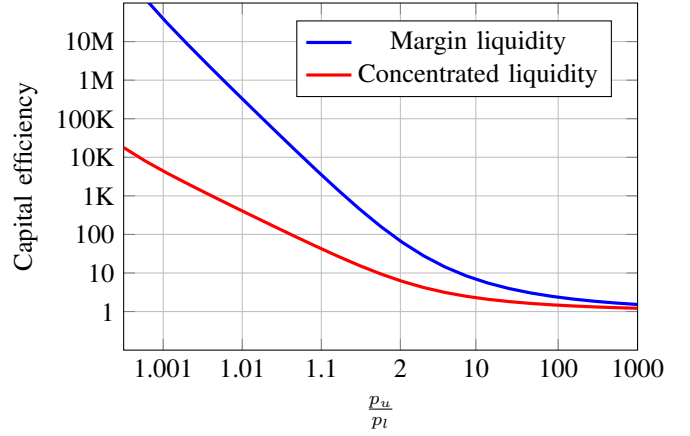

%% file: text/08conclusion.tex
Liquidity is one of the most crucial factors when traders choose an exchange, regardless of whether the exchange operates in a centralized or decentralized manner. However, the inherent risk of divergence loss in DEX-AMM hinders the market participants from providing enough liquidity. In this regard, we introduce a new concept of liquidity, \emph{margin liquidity}, and propose a highly profitable liquidity-providing position, which can be applied to both CEX and DEX, to attract risk-taking market participants. Also, we extend this position to hedge the risk of divergence loss of LPs so that LPs can provide more liquidity. Furthermore, we introduce a representative strategy for market participants who want to provide liquidity with our proposed position. Our representative strategy outperforms a simple holding baseline with a higher ROR and a lower MDD in the backtest on the historical data of the ETH/BTC market. Also, we show that our margin liquidity is 8K$\times$ more capital efficient than the concentrated liquidity proposed by Uniswap V3.